# Antiferromagnetic Kitaev interaction in $J_{eff}$=1/2 cobalt honeycomb materials $Na_3Co_2SbO_6$ and $Na_2Co_2TeO_6$

Chaebin Kim[1,2], Jaehong Jeong[2,3*], Gaoting Lin[4], Pyeongjae Park[1,2], Takatsugu Masuda[5], Shinichiro Asai[5], Shinichi Itoh[6], Heung-Sik Kim[7], Haidong Zhou[8], Jie Ma[4,9#], and Je-Geun Park[1,2$]

[1]Center for Quantum Materials, Seoul National University, Seoul 08826, Korea

[2]Department of Physics and Astronomy, Seoul National University, Seoul 08826, Korea

[3]Center for Correlated Electron Systems, Institute for Basic Science, Seoul 08826, Korea

[4]Key Laboratory of Artificial Structures and Quantum Control, School of Physics and Astronomy, Shanghai Jiao Tong University, Shanghai 200240, China

[5]Institute for Solid State Physics, The University of Tokyo, Chiba 277-8581, Japan

[6]Institute of Materials Structure Science, High Energy Accelerator Research Organization, Tsukuba 305-0801, Japan

[7]Department of Physics, Kangwon National University, Chuncheon 24311, Korea

[8]Department of Physics and Astronomy, University of Tennessee, Knoxville, Tennessee 37996, USA

[9]Shenyang National Laboratory for Materials Science, Institute of Metal Research, Chinese Academy of Sciences, 110016 Shenyang, China.

*hoho4@snu.ac.kr

#jma@sjtu.edu.cn

$jgpark10@snu.ac.kr

**Abstract**

Finding new materials with antiferromagnetic (AFM) Kitaev interaction is an urgent issue for quantum magnetism research. We conclude that $Na_3Co_2SbO_6$ and $Na_2Co_2TeO_6$ are new honeycomb cobalt-based systems with AFM Kitaev interaction by carrying out inelastic neutron scattering experiments and subsequent analysis. The spin-orbit excitons observed at 20-28 meV in both compounds strongly support the idea that $Co^{2+}$ ions of both compounds have a spin-orbital entangled $J_{eff}$=1/2 state. Furthermore, we found that a generalized Kitaev-Heisenberg Hamiltonian can describe the spin-wave excitations of both compounds with additional 3[rd] nearest-neighbor interaction. Our best-fit parameters show significant AFM Kitaev terms and off-diagonal symmetric anisotropy terms of a similar magnitude in both compounds. We also found a strong magnon-damping effect at the higher energy part of the spin waves, entirely consistent with observations in other Kitaev magnets. Our work suggests $Na_3Co_2SbO_6$ and $Na_2Co_2TeO_6$ as rare examples of the AFM Kitaev magnets based on the systematic studies of the spin waves and analysis.

**Keywords:** Kitaev model, Magnetism, Spin-orbit entanglement, Quantum Spin Liquid

# 1.    Introduction

The quantum spin liquid phase has attracted tremendous attention over the past decades or so. It has since become one of the most active topics in condensed matter physics since the exact solution of the Kitaev model was demonstrated to host a new type of so-called Kitaev quantum spin liquid (KQSL) [1–3]. The Kitaev model, an exactly solvable quantum spin ($S$=1/2) model, is well-defined on a honeycomb lattice with an Ising-type interaction between the nearest-neighbors, of which the Ising axis is bond-dependent (see Fig. 1a) [1]. The orthogonality of three different Ising axes forces competition among the bond-dependent interactions, leading to magnetic frustration. This exchange frustration gives rise to infinite degeneracy in the ground state, leading to the KQSL, which can be recast using the Majorana fermions operators [1].

The entanglement of spin and orbital sectors is essential to realize such bond-dependent anisotropic interaction. Thus, most effort has been so far focused on the $4d$- and $5d$-electron systems with a strong spin-orbit coupling (SOC) [4,5]. For example, $A_2IrO_3$ [6–9] and α-RuCl$_3$ [10–19] have been suggested to show a large ferromagnetic (FM) Kitaev interaction. However, the realization of KQSL remains elusive as a long-range magnetic order invariably kicks in at higher temperatures for all those candidates due to Heisenberg interactions of nonnegligible magnitude. Another interesting and equally important point is that no antiferromagnetic (AFM) Kitaev system has been reported so far, although several systems with the FM Kitaev interaction have been found experimentally [3]. Closely related to the theme of this paper, the latest theories [20–24] predict that AFM Kitaev materials can host two different classes of KQSL. One is associated with the usual Z$_2$ gauge field and another with the $U$(1) gauge field, whereas FM Kitaev systems can, in theory, have only the former. Therefore, it adds further importance to finding AFM Kitaev materials.

It has recently been suggested that, despite a much smaller SOC, a significant Kitaev interaction can be realized in 3$d^7$ Co$^{2+}$ systems, too [25–27]. Interestingly, it was found that even Cu$^{2+}$ can exhibit such $J$-physics under certain conditions [28,29]. The $d^7$ electrons of Co$^{2+}$ with a $(t_{2g})^5(e_g)^2$ configuration in an octahedral crystal field can, in principle, possess a spin-orbital entangled (SOE) multiplet state with $S$=3/2 and $L_{eff}$=1, as shown in Fig. 1b. Several features make the $d^7$ systems unique as compared to the $d^5$ systems. First of all, the spin-active $e_g$ electrons are expected to play a significant role in the exchange process for the $d^7$ systems [25–27], while only the $t_{2g}$-$t_{2g}$ channel plays a role in the $d^5$ systems. Those new contributions to the non-Kitaev exchange interactions can essentially cancel each other, and even larger Kitaev coupling can be expected for the $t_{2g}$-$e_g$ process [25–27]. Therefore, this new and intriguing exchange process makes the honeycomb cobaltates a promising new candidate to realize the ideal Kitaev model [27].

Honeycomb-layered cobaltates $Na_3Co_2SbO_6$ (NCSO) and $Na_2Co_2TeO_6$ (NCTO), theoretically proposed KQSL candidates [27], have a similar atomic structure with a honeycomb layer of edge-sharing $CoO_6$ octahedra; $SbO_6$ and $TeO_6$ octahedra are located at the honeycomb center, respectively [30–36]. Both compounds possess a common zig-zag magnetic ordering: NCSO has $T_N$=8 K with a propagation vector $\mathbf{k}$=(1/2, 1/2, 0) while NCTO has $T_N$=27 K with $\mathbf{k}$=(1/2, 0, 0) [31–33,36]. To test the theoretical proposal of KQSL, it is now imperative to determine the strength and the sign of Kitaev coupling and non-Kitaev interactions experimentally. This information will eventually lead to further highly original ways of achieving the KQSL phase by tuning external parameters.

We report the magnetic excitations of NCSO and NCTO using powder inelastic neutron scattering (INS) experiments. Our data and subsequent analysis provide convincing evidence for a significant AFM Kitaev interaction and Heisenberg and off-diagonal anisotropy exchanges of comparable magnitude for both compounds, which is crucial to achieving KQSL. Our observation of the corresponding spin-orbit (SO) exciton confirms the SOE $J_{eff}$=1/2 ground state in both compounds, which is the main ingredient to realize the Kitaev interaction. In contrast with the original theoretical prediction in Ref.[27], however, our analysis with detailed model calculations concludes that both compounds have an AFM Kitaev coupling, not FM Kitaev coupling as initially predicted. We also find large magnon decays for spin waves at higher energies predicted to arise from the two-magnon process.

## 2. Methods

Polycrystalline NCTO and NCSO samples were prepared by a conventional solid-state reaction method with the recipe provided in Ref. [30]. INS experiments were performed at the HRC beamline of J-PARC, Japan [37]. The data were collected at the following conditions: for NCSO at $T$=3, 15, and 50 K with the fixed incident neutron energies of $E_i$=7.1, 12.19, 16.54, 35.61, 50.9, and 122.6 meV; for NCTO at $T$=3, 10, 20, 30, 50, and 95 K with $E_i$=11.44, 16.54, 50.9 and 122.6 meV. The measured data were reduced and binned using the MSLICE program of the DAVE suite [38]. The spin-wave spectra and INS cross-section were calculated using the SpinW library [39]. We also developed our own code for calculating two-magnon spectra.

## 3. Results and Analysis

### 3.1 Spin-orbit exciton

The SO exciton between the $J_{eff}$= 1/2 ground state and the excited states is commonly observed among cobalt compounds supporting the SOE state for $Co^{2+}$ ions [40–44]. Since a trigonal distortion of octahedra further splits the $J_{eff}$=3/2 state into two levels, the lowest exciton energy is determined by a combination of SOC ($\lambda$) and trigonal crystal field ($\Delta$) [27]. Figure 2 shows the temperature dependence of the SO excitons in the data measured of NCSO and NCTO with $E_i$=122.6 meV. For NCSO (Fig. 2a-c), the exciton is observed at 28.1 meV above $T_N$, and its position slightly increases by 1.1 meV below $T_N$. For NCTO (Fig. 2d-f), it is located at 23.1 meV and moves to 25.3 meV upon entering the magnetic ordering. This change in the exciton energy through $T_N$ can be interpreted as the Zeeman splitting due to a molecular magnetic field induced by the magnetic ordering (Fig. 2g).

The exciton energy can be determined using the data taken with $E_i$=50 and 100 meV (see Fig 3). To understand the transition of this crystal-field excitation accurately, we used a single-ion Hamiltonian given as

$$H = H_{SO} + H_{tri} + H_{MF} = \lambda \boldsymbol{L} \cdot \boldsymbol{S} + \Delta \left( L_{\hat{n}}^2 - \frac{2}{3} \right) + h_{mf} S_{\hat{b}} \, ,$$

where $\lambda$ is the spin-orbit coupling, $\Delta$ is the trigonal crystal field, and $h_{mf}$ denotes the molecular field from the magnetic ordering. Note that the $\hat{n}$ vector is given along the [1 1 1] direction for the local frame. These parameters in the Hamiltonian can be experimentally determined by measuring several crystal field excitations, as shown in Fig. 3. Interestingly, a magnetic ordering produces a shift in the transition energy through a molecular magnetic field, which shows a clear difference depending on the sign of Δ. For instance, the lowest spin-orbit exciton energy is seen to increase for Δ>0 with the magnetic ordering. On the other hand, it splits into two modes for Δ<0, and the lower one moves slightly towards lower energies (see Fig. 2h). As the energy shift is positive in our data, we can conclude that both our samples have a positive sign of trigonal distortion. The best-fitting results of the observed energy change can be obtained with λ=25 meV, Δ=12 meV, $h_{mf}$=0.4 meV for NCSO and λ=21 meV, Δ=13 meV, $h_{mf}$=0.6 meV for NCTO.

### 3.2 Spin-wave spectrum and its analysis

Figures 4a-b show the magnon spectra of NCSO and NCTO, respectively, measured at $T$=3 K with an incident neutron energy of $E_i$=16.54 meV. To correct the low-temperature data for background and phonon contaminations, we used the high-temperature data measured well above $T_N$: at 50 K for NCSO and at 95 K for NCTO. Despite having similar atomic and magnetic structures, the magnon dispersions show strikingly distinctive features of NCSO and NCTO. For NCSO, a strong upturn-shaped dispersion is observed at low $Q$<1 Å$^{-1}$ and $E$~1-3 meV with a small gap of 0.6 meV, while the higher energy data show a weak arch-shaped dispersion persisting up to 8 meV. On the other hand, NCTO shows a dispersionless excitation at ~7 meV and strong triangular-shaped dispersions below ~3 meV with a gap of 0.4 meV.

To explain the observed magnon spectra, we used the generalized Kitaev-Heisenberg (GKH) pseudospin $\tilde{S} = 1/2$ Hamiltonian:

$$
\begin{aligned}
H &= \sum_{n=1,3} J_n \sum_{<i,j>_n} \tilde{\boldsymbol{S}}_{\boldsymbol{i}} \cdot \tilde{\boldsymbol{S}}_{\boldsymbol{j}} \\
&+ \sum_{<i,j> \in \alpha\beta(\gamma)} [K\tilde{S}_i^\gamma \tilde{S}_j^\gamma + \Gamma(\tilde{S}_i^\alpha \tilde{S}_j^\beta + \tilde{S}_i^\beta \tilde{S}_j^\alpha) + \Gamma'(\tilde{S}_i^\alpha \tilde{S}_j^\gamma + \tilde{S}_i^\gamma \tilde{S}_j^\alpha + \tilde{S}_i^\beta \tilde{S}_j^\gamma + \tilde{S}_i^\gamma \tilde{S}_j^\beta)], \quad \dots (1)
\end{aligned}
$$

where $J_n$ is a Heisenberg coupling between the $n$th nearest neighbors, $K$ is a Kitaev interaction, and Γ/Γ' denotes a symmetric anisotropy (off-diagonal) exchange interaction.

For each bond, we can distinguish an Ising axis $\gamma$, labeling the bond $\alpha\beta(\gamma)$, where $\alpha$ and $\beta$ are the other two remaining axes. Since the 2$^{nd}$ nearest-neighbor Heisenberg interaction is known to be relatively small in many honeycomb compounds, we only considered the 1$^{st}$ and 3$^{rd}$ nearest-neighbor Heisenberg interactions in our analysis.

We also used a simple anisotropic Heisenberg (XXZ) Hamiltonian for a fair comparison with the generalized Kitaev-Heisenberg model. Note that the XXZ model is commonly used to explain the spin dynamics of cobalt honeycomb compounds [45,46]. We added additional single-ion anisotropy to the XXZ model to align spins orthogonal to the propagation vector consistent with the reported magnetic structure and spin gap. The XXZ model used in this study is written as

$$\mathrm{H} = \sum_{n=1,3} J_n \sum_{<i,j>_n} \left[ S_i^x S_j^x + S_i^y S_j^y + \alpha S_i^z S_j^z \right] + D \sum_i (\hat{e} \cdot S_i)^2 \ ,$$

where $\alpha \in [0, 1]$ is the spin anisotropy parameter, $D$ is the strength of single-ion anisotropy, and $J_n$ is the Heisenberg exchange interaction with the first and the third nearest neighbors. $\hat{e}$ is a unit vector orthogonal to the propagation vector. We used the following best-fitting parameters: for NCSO $J_1$=-3.6, $J_3$=1.9, $\alpha$=0.8, and $D$=-0.7 meV; for NCTO $J_1$=-2.1, $J_3$=2.1, $\alpha$=0.95, and $D$=-0.1 meV.

Interestingly enough, we obtained the best-fitting parameters with a significant AFM Kitaev coupling by searching the vast parameter space, summarized in Table 1. We calculated the powder-averaged magnon spectra using Eq. (1) with the best-fitting parameters and show the final results in Fig. 4c-d for NCSO and NCTO, respectively. We also obtained the best-fitting parameters with the anisotropic Heisenberg (XXZ) Hamiltonian with a single-ion anisotropy term, supporting the reported magnetic structure and the magnon gap in Fig. 4e-f. We made a detailed comparison between the two models by employing const-$Q$ cuts, integrated over the $Q$ range as denoted with vertical boxes in Fig. 4a-d. As shown in Fig. 4g-j, the GKH model provides the best agreement with an AFM Kitaev coupling of a few meV. The simpler XXZ model reproduces some of the brief shapes of the measured dispersions, with much less consistency about the detailed features. We thus conclude that the GKH model is the best one for both systems.

Following the predictions of the recent theoretical reports [25–27], we also examined the FM Kitaev ($K$<0) coupling as a possible alternative model for the observed spin-waves spectra [25–27]. Figure 5 shows the calculated powder-averaged spin-waves spectra and the optimized magnetic structures with the best-fitting FM Kitaev parameters: for NCSO, $J_1$=-2.1, $J_3$=1.2, $K$=-4, $\Gamma$=-0.7, and $\Gamma'$=0.6 meV; for NCTO, $J_1$=-0.1, $J_3$=1.4, $K$=-7.4, $\Gamma$=-0.1, and $\Gamma'$=0.05 meV. We note that the FM Kitaev model appears to show a similar degree of agreement regarding the AFM Kitaev model with the data. However, the magnetic structures optimized for each model within the spin-waves calculations differ from each other: the significant difference is about the direction of magnetic moments. For example, the AFM Kitaev model predicts moments aligned orthogonal to the propagation vector, whereas it is parallel with the propagation vector for the FM Kitaev coupling. Unfortunately, any of those optimized structures of both AFM and FM Kitaev models does not precisely match the reported ones (left in Fig. 5). However, we conclude that the magnetic structures with the AFM Kitaev model are much closer to the reported ones of NCTO. At the same time, there is no way for the dominant FM Kitaev model to become a magnetic moment lying on the *ab*-plane (see details in the discussion section). We note that the optimized magnetic structure with the AFM Kitaev model for NCTO has an additional canting along the *c*-axis, absent in the reported magnetic structure. In contrast, we confirm that our optimized magnetic structure for NCSO shows excellent agreement with the reported single-crystal neutron diffraction data [36]. As a passing comment, we would like to note that even for the AFM Kitaev model, there can be two possibilities: one is our robust AFM model, and another is a small AFM model, as discussed in ref. [47]. As we demonstrated in this paper, the robust AFM model seems to be a better fit for the inelastic neutron scattering data. It is of our current view that we need high-resolution inelastic neutron scattering data collected on single crystals samples to distinguish among the several contending models.

### 3.3   Magnon damping effect

Although our AFM Kitaev model reproduces the main features of the measured magnon spectra, it is seen to overestimate the intensity at high energies. We note that similar damped magnon dispersions were also observed at high energies in α-RuCl$_3$ [48–50]. A well-accepted view is that such a significant damping effect can originate from a two-magnon process and the renormalization effect of the Kitaev interaction [48–50]. To test this scenario qualitatively, we calculated the two-magnon density of state (DOS) using our home-built code customized for the GKH model with the best-fitting parameters.

To examine the magnon damping effect in our data, we first calculate the non-interacting two-magnon density of state (DOS) with

$$D(\mathbf{q}, E) = \frac{1}{N} \sum_{i,j} \sum_{\mathbf{k}} \delta(E - E_{\mathbf{k},i} - E_{\mathbf{q-k},j}),$$

where $\mathbf{k}$ is a set of $\mathbf{q}$ points on the equally spaced mesh in the 1$^{st}$ Brillouin zone, $E_{\mathbf{k},i}$ is the $i$th magnon's energy dispersion, and N is a normalization factor. We developed our own code to calculate the two-magnon spectra following the method presented in Ref.[51]. In the magnon decay process, $D(\mathbf{q}, E_{\mathbf{q}})$ is the number of possible decay channels, with a single magnon at ($\mathbf{q}, E_{\mathbf{q}}$) decaying into two magnons with the kinematic constraint of $E_{\mathbf{q}} = E_k - E_{k-q}$. To a first order of approximation, $D(\mathbf{q}, E_{\mathbf{q}})$ is known to give a good estimate for the damping[48–50]. Figure 6 plots the calculated two-magnon DOS for NCSO and NCTO along the high symmetry lines. We note that for NCSO, the strong two-magnon DOS overlaps all over the upper modes of single magnons, which can describe well the highly damped high-energy spectra at 4-8 meV in our data. In comparison, the two-magnon DOS is present on the 5-6 meV, slightly below the flat spectra near 7 meV for NCTO. Based on these results, we suggest that the high-energy damping seen in the high-energy magnon spectra originates from the two-magnon decay process. Similar observations have been made for systems with a strong anisotropic exchange, such as the Kitaev coupling and off-diagonal symmetric anisotropy terms [48–50].

## 4.    Discussion

We now like to have a more in-depth discussion on the two AFM and FM model Hamiltonians from the standpoint of theory and experiment. First of all, we note that though the recent theoretical works predicted only an FM Kitaev coupling in these materials, a small perturbation in the local environment is also known to significantly affect the hopping process and the resulting exchange parameters [52]. Especially in the $d^7$ systems, the exchange path can be even more complicated due to the spin-active $e_g$ electrons. Thus, we believe that our choice of AFM Kitaev coupling may well be justified from this theoretical viewpoint.

## 4.1 Magnetic phase diagram with K-Γ-Γ' model

We performed matrix diagonalization with both FM and AFM Kitaev interaction with other off-diagonal anisotropies to further examine the correlation between magnetic moment and the anisotropic exchange interactions. In the generalized Kitaev-Heisenberg model with zig-zag magnetic order, the spin direction is solely determined by the Kitaev term and other off-symmetric anisotropy; which corresponds to the eigenvector of the matrix M: [53]

$$M = \begin{pmatrix} 2K & -\Gamma + 2\Gamma^{'} & \Gamma \\ -\Gamma + 2\Gamma^{'} & 2K & \Gamma \\ \Gamma & \Gamma & 0 \end{pmatrix}$$

With diagonalization, we can get the eigenvalues

$$E_p = \Gamma - 2\Gamma^{'} + 2K$$

$$E_{\pm} = \Gamma^{'} - \frac{\Gamma}{2} + K \pm \frac{A}{2}$$

where $A = \sqrt{9\Gamma^2 + 4\Gamma^2 + 4K^2 - 4\Gamma\Gamma^{'} - 4K\Gamma + 8K\Gamma^{'}}$

The eigenfunction of each eigenvalue is given as below in the monoclinic frame.

$$v_p = (0,1,0)$$

$$v_{\pm} = \left( -\frac{\sqrt{2}}{8} \frac{7\Gamma + 2\Gamma^{'} + 2K \pm 3A}{\Gamma^{'} - \Gamma + K}, 0, 1 \right)$$

If $E_p$ is the lowest energy, spin aligns parallel to the bond direction and lies on the ab-plane. On the other hand, for the $E_{\pm}$ case, spin aligns orthogonal to the bond direction with two possibilities: either lying on the ab-plane or canting to the c-axis (see Fig. 7).

We particularly examined in Figure 8 how the angle of the moment with respect to the honeycomb plane varies in the parameter space of the K-Γ-Γ′ model. The white area in Fig. 8 indicates the magnetic structure with moments parallel to the bond direction. Furthermore, the colored area represents the magnetic structure with spin orthogonal to the bond direction. As one can see, the in-plane moment configuration can easily be accessed with AFM Kitaev interaction. On the other hand, the magnetic moment is mostly canted to the $c$-axis with FM Kitaev interaction. Since the reported spin configurations are almost lying on the $ab$ plane, this opposite tendency from the sign of Kitaev interaction indicates that the AFM Kitaev model is more appropriate for explaining the spin-wave and magnetic structure of NCSO and NCTO. The relaxed magnetic structure for each fitted model is located in the phase space of Figure 8. Note that although the magnetic moment's angle of our AFM Kitaev model and M. Songvilay's FM Kitaev model [54] is similar, the sign of the angle is different. Another worthy note is that a recent study [55] suggests that the magnetic structure of NCTO might be triple-Q zig-zag order rather than a simple single-Q zig-zag order with three equivalent domains. Such conflicting reports about the magnetic ground state of NCTO [31,32,55] indicate that it needs reinvestigation to define the more reliable magnetic Hamiltonian.

## 4.2 Comparison with single-crystal data

To further compare the spin waves of different models, we have calculated the magnon dispersion curves of NCSO and NCTO with each reported parameter. We also considered how the possible domain structures might affect the measured spin waves. Figure 9 shows the spin-wave spectrum of NCSO and NCTO along the high-symmetric line at the Brillouin zone with different models with three equivalent magnetic domains highlighted by different colors (see Fig. 9). For the NCTO, the striking feature of the AFM Kitaev model is no crossing band between lower and higher magnon branches. Other models display such crossing through the high symmetric lines.

Interestingly, the spin wave's gapped-like feature appears differently in NCSO. In the case of NCSO with the AFM Kitaev model, the two magnon parts are very close to each other, whereas the branches are gapped at the FM Kitaev models. This gapped-like feature will give us insight and help us choose the correct magnetic Hamiltonian when measuring the single-crystal data.

And we would like to specifically examine the low-energy spin-wave dispersion for several models as the experimental results were reported [55]. In particular, we want to examine a possible triple-Q zig-zag structure. In the figures, we simulated the spin-wave spectrum along the same direction as the data. Figure 10 shows the calculation of spin-wave dispersion along the (H,0) direction. Our parameter set from powder measurement shows sound agreement with the reported single-crystal data. While the full-dispersion of spin-wave data for detailed comparison is needed, our simulation implies that the simple zig-zag order can explain the observed spin-wave. Moreover, the calculated spin-wave spectrum from our AFM Kitaev model is in better agreement with the data than other reported models for NCTO [54,55].

## 5.    Conclusion

In summary, we measured the magnetic excitations in two Co-based KQSL candidates, NCSO and NCTO, using inelastic neutron scattering. We observed the temperature-dependent SO exciton in both compounds, explained by the crystal field excitation between the SOE $J_{eff}=1/2$ ground state and $J_{eff}=3/2$ excited state with a positive trigonal distortion. We determined a considerable AFM Kitaev interaction in the magnon spectra with a comparable size of Heisenberg and off-diagonal symmetric anisotropy exchange interactions for the GKH model. We also found the strong magnon decay over a wide range of high-energy spectra, which can be interpreted as the two-magnon process enhanced by anisotropic exchanges. This work provides experimental evidence for the AFM Kitaev interaction in real materials and opens up new opportunities to achieve KQSL in real materials.

**Acknowledgments**

The authors would like to thank H. B. Cao and M. A. McGuire for private communications [56] on the magnetic structure of NCSO and S. D. Ji for helpful discussion. The work at CQM and SNU was supported by the Leading Researcher Program of the National Research Foundation of Korea (Grand No. 2020R1A3B2079375). The INS experiment was performed at the MLF of J-PARC under a user program (Proposal No. 2019B0350).

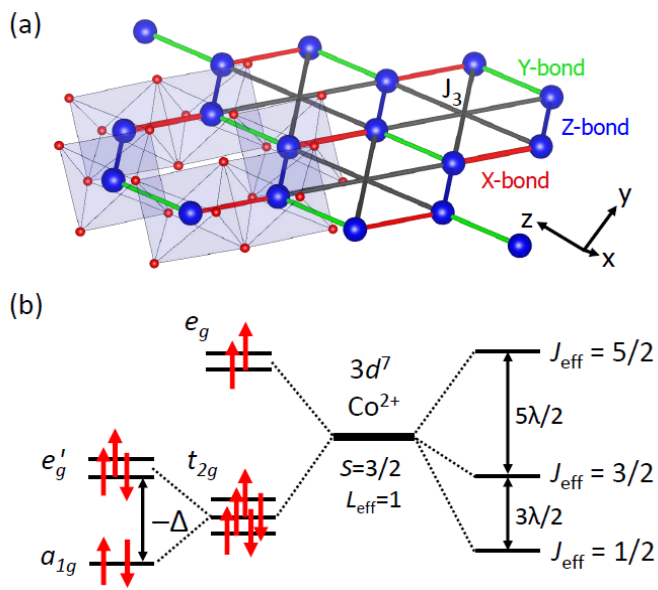

Figure 1. (a) Schematic view for Kitaev interaction on a honeycomb lattice. Each red, green, and blue line indicates the 1$^{st}$ nearest-neighbor interactions associated with the local x, y, z axes in the Kitaev model. Grey lines indicate the additional 3$^{rd}$ nearest-neighbor interaction. (b) Splitting of the degenerate $d^7$ states due to an octahedral and trigonal crystal field in a single-electron picture (left diagram) and the spin-orbit coupling in a multi-electron picture (the right-sided diagram).

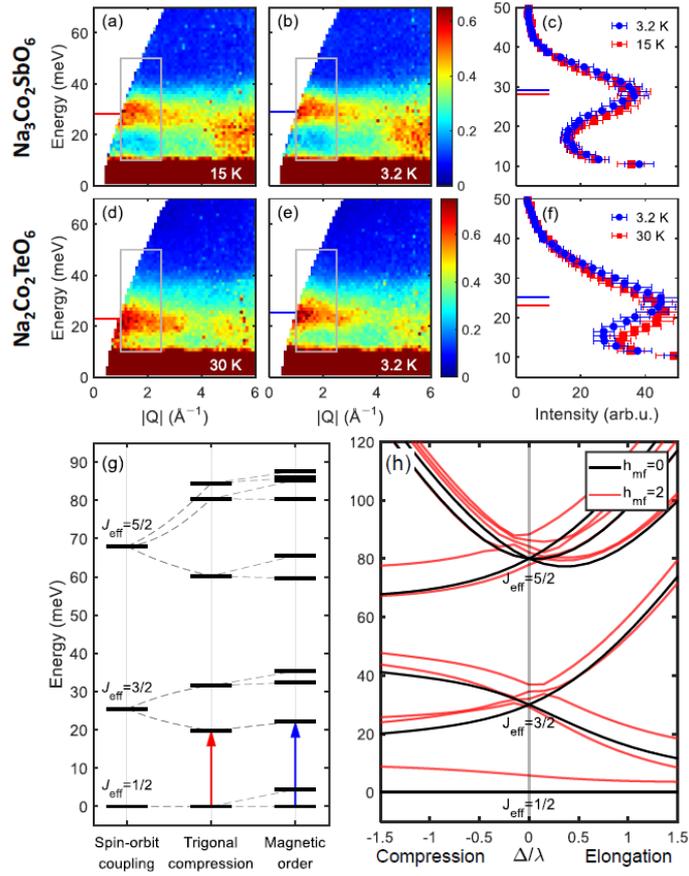

Figure 2. Temperature dependence of the spin-orbit excitons in (a-c) $Na_3Co_2SbO_6$ and (d-f) $Na_2Co_2TeO_6$. Grey boxes in (a,b,d,e) denote the integration range, $Q$=[1, 2.5] Å$^{-1}$, for constant-$Q$ cuts in (c,f). (g) Splitting of the spin-orbital entangled states due to a positive trigonal crystal field and molecular magnetic field from magnetic ordering. (h) Splitting of crystal field levels due to a trigonal crystal field (black) and further splitting due to a molecular magnetic field induced by magnetic ordering (red).

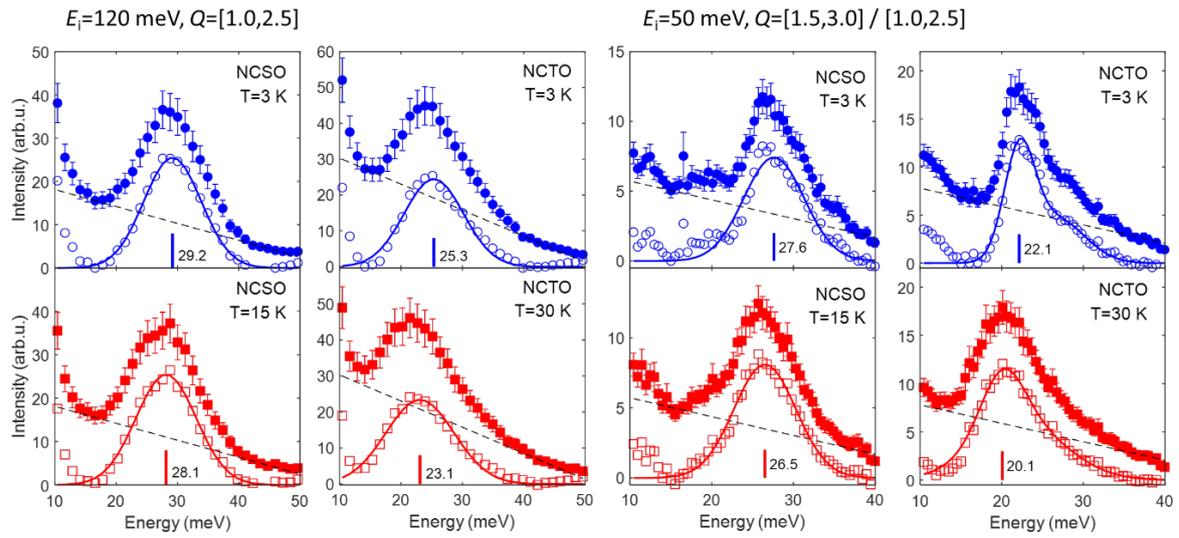

Figure 3. Fitting of spin-orbit excitons. The obtained fitting parameters were used for the subsequent crystal field analysis.

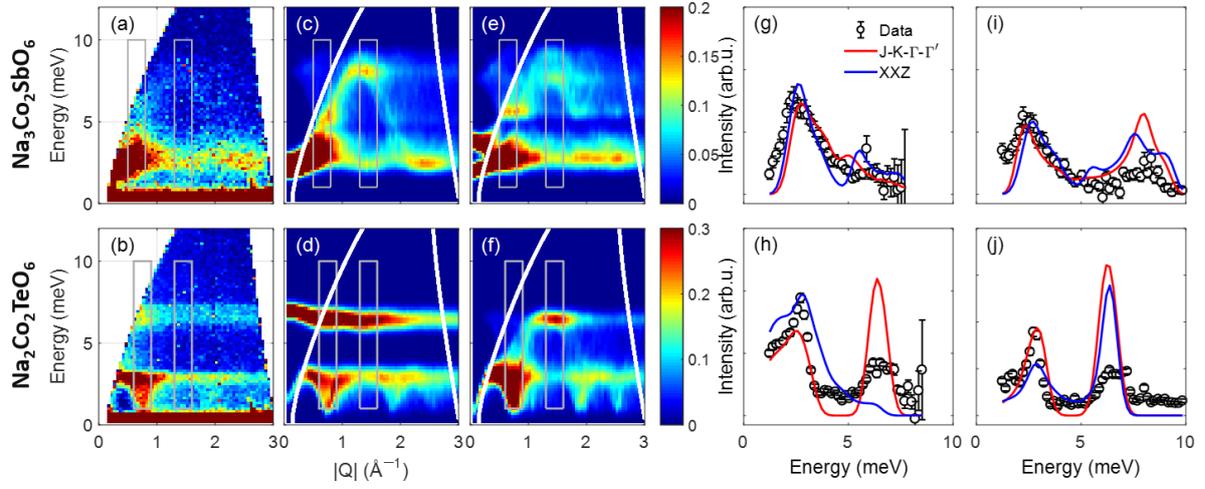

Figure 4. (a,b) Magnon spectra of NCSO and NCTO measured at $T$=3.2 K with $E_i$=16.54 meV. Calculated powder magnon spectra (c,d) using the generalized Kitaev-Heisenberg model and (e,f) using the XXZ model with the best-agreement parameters. Comparison of constant-$Q$ cuts, (g,i) integrated over $Q$=[0.5 0.8] and [1.3 1.6] Å$^{-1}$ for NCSO and (h,j) integrated over $Q$=[0.6 0.9] and [1.3 1.6] Å$^{-1}$ for NCTO.

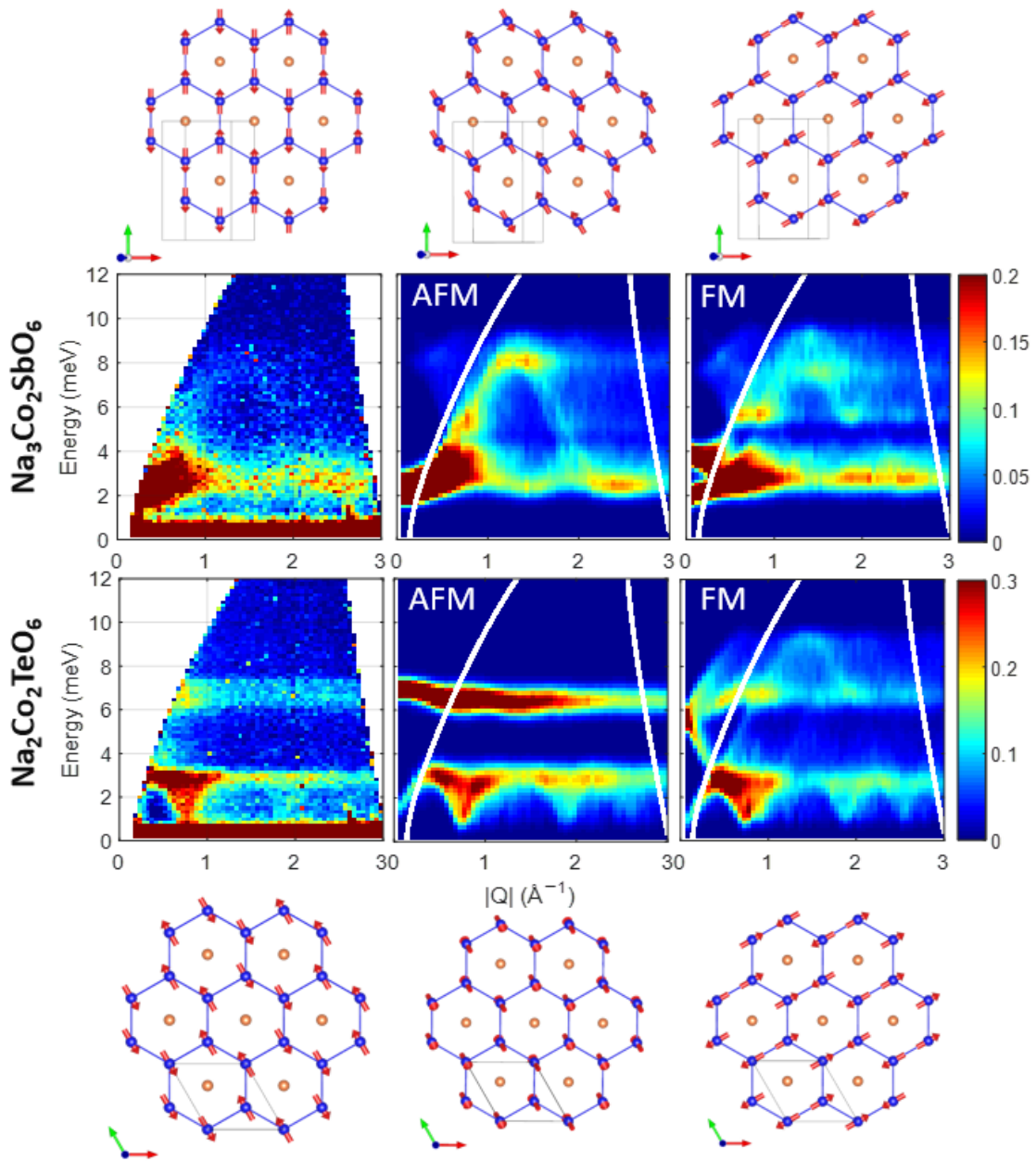

Figure 5. Spin-wave spectra measured at $T$=3 K (left) and powder-averaged spectra calculated with AFM Kitaev model (center) and FM Kitaev model (right). The reported magnetic structures (left) and model-optimized magnetic structures are plotted together.

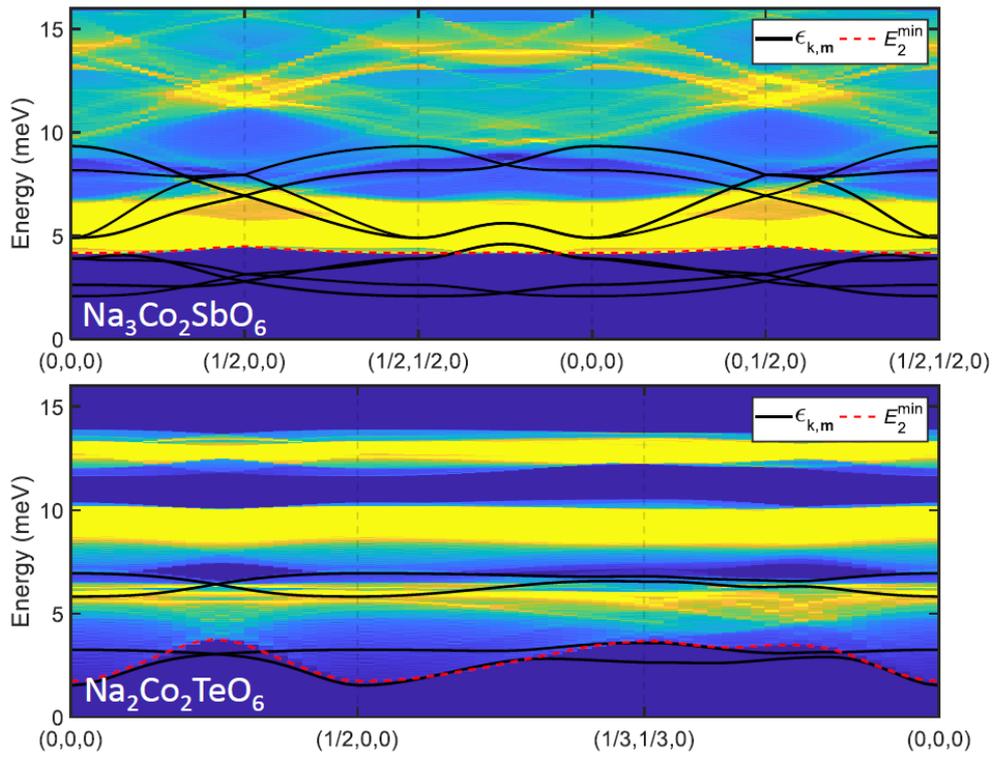

Figure 6. Two-magnon DOS calculation for both compounds. The black line in each figure shows the linear spin-wave dispersion $\varepsilon_{\mathbf{k},m}$, and intensity indicates the number of two-magnon DOS. The dashed red line indicates the lower bound of the two-magnon continuum $E_2^{min}$.

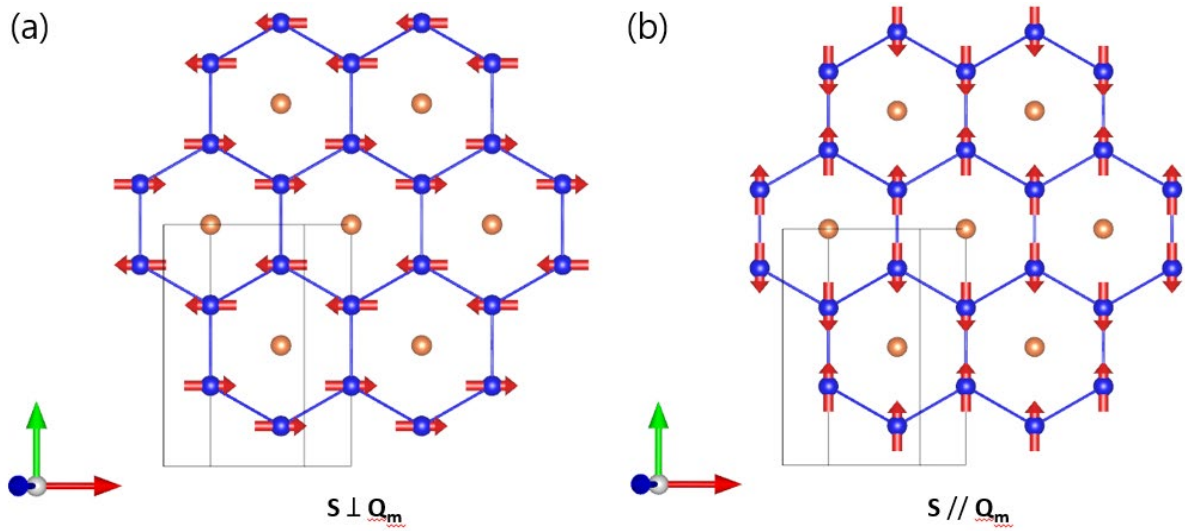

Figure 7. Two possible magnetic structures from the generalized Kitaev-Heisenberg model with zig-zag order. (a) The eigenstate of $E_\pm$. The spin aligned orthogonal to the bond direction. (b) The eigenstate of $E_p$. The spins are aligned parallel to the bond direction. The propagation vector is set to $Q = (0, 1, 0)$ in this figure.

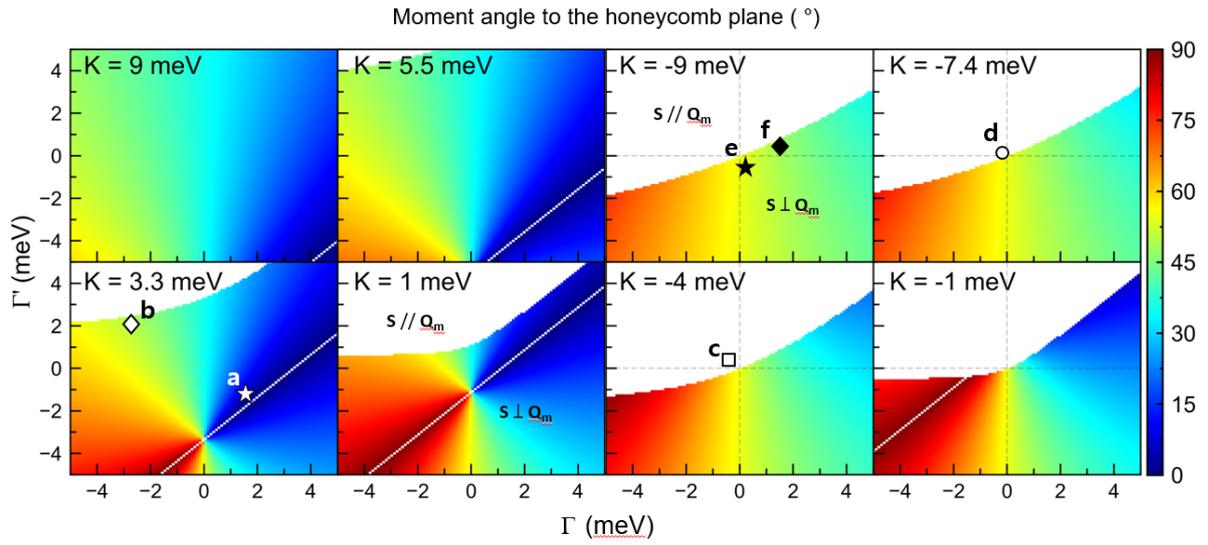

Figure 8. Phase diagram of the moment angle with respect to the honeycomb plane in the K-Γ-Γ′ parameter space. (a-b) are the locations of our best-fitting parameter set of NCSO and NCTO with the AFM Kitaev model. (c-d) are the location of our best-fitting parameter set of NCSO and NCTO with the FM Kitaev model. (e-f) are the location of M. Songvilay's parameter set of NCSO and NCTO, respectively[54].

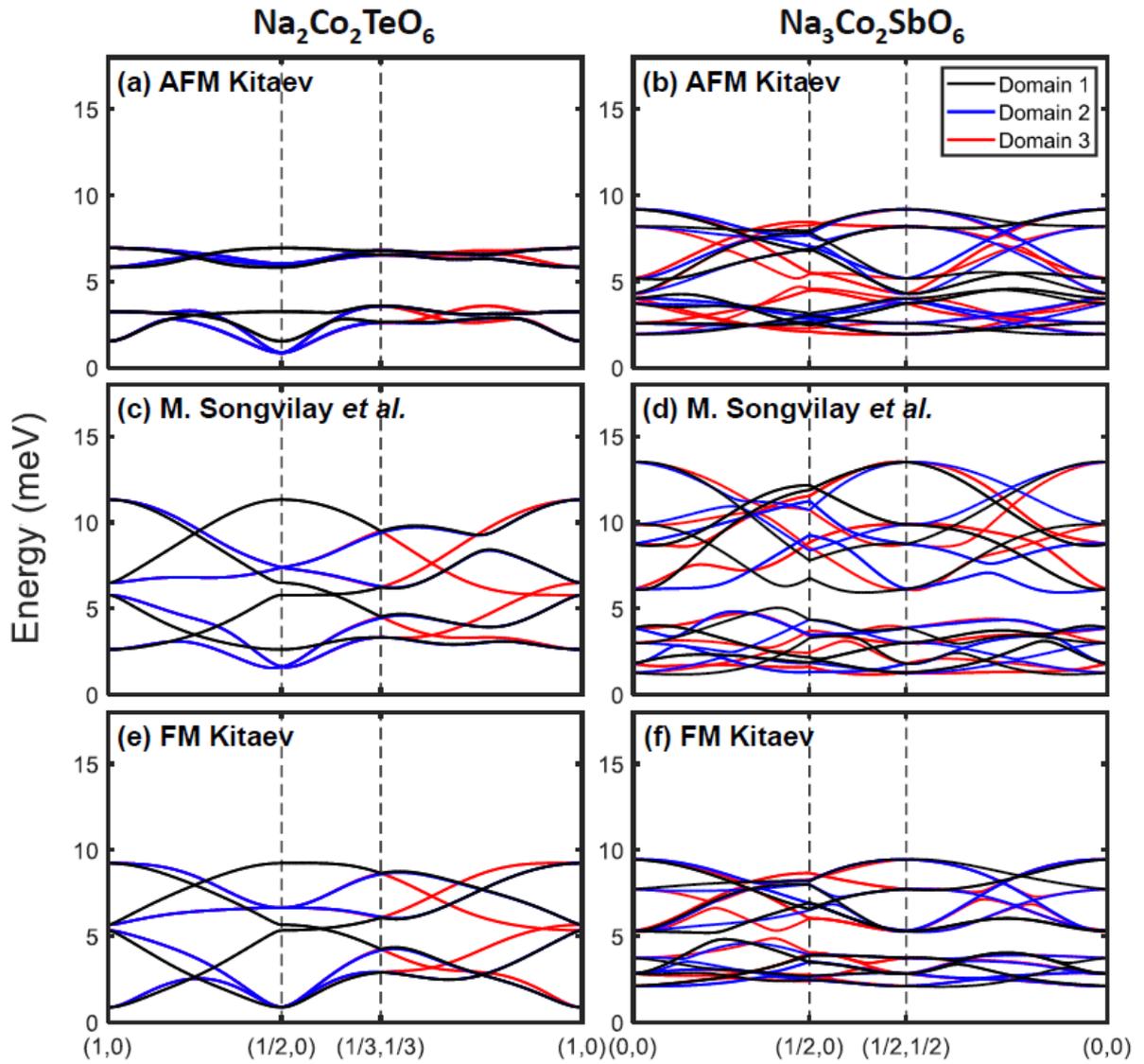

Figure 9. (a-f) The spin-wave dispersion of NCSO and NCTO along the high-symmetry lines at the Brillouin zone. (a-b) GKH model with dominant AFM Kitaev interaction from our fitted parameter. (c-d) GKH model with dominant FM Kitaev interaction from Ref. [54]. (e-f) GKH model with dominant FM Kitaev interaction from our fitted parameter.

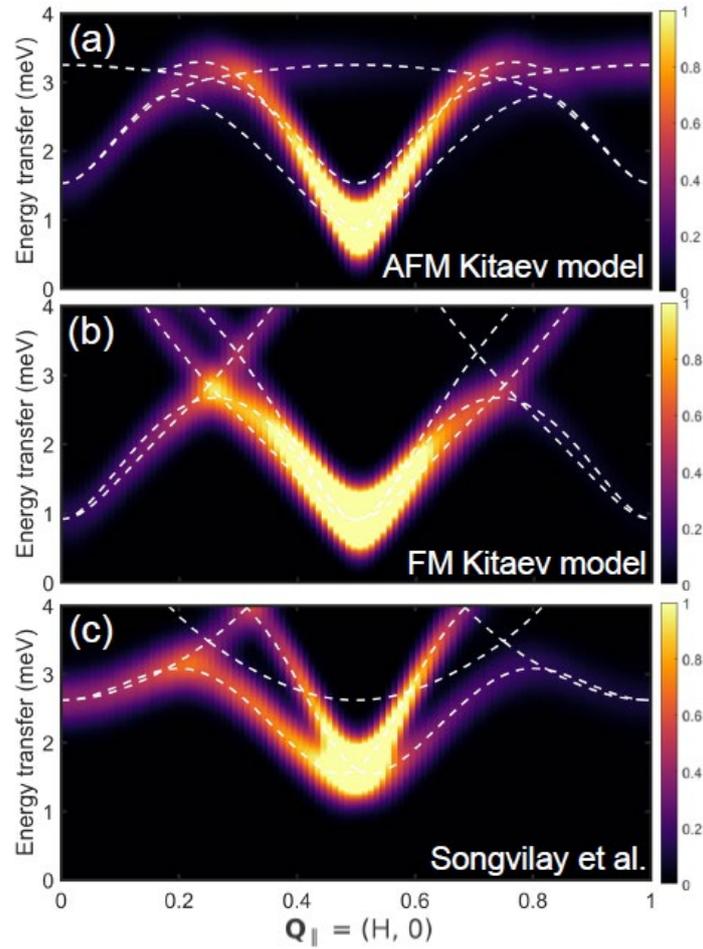

Figure 10. (a-c) Calculated spin-wave dispersions and its domain-averaged dynamical structure factors of GKH models from our works and Songvilay et al. [54]. The white lines indicate the spin-wave dispersion for each model, considering three equivalent domains.

Table 1. The best-fitting parameters with the generalized Kitaev-Heisenberg model.

| | $J_1$ (meV) | $J_3$ (meV) | $K$ (meV) | $\Gamma$ (meV) | $\Gamma'$ (meV) |
|---|---|---|---|---|---|
| NCSO | -4.70(5) | 0.95(1) | 3.60(4) | 1.30(5) | -1.40(4) |
| NCTO | -1.50(5) | 1.50(2) | 3.30(10) | -2.80(5) | 2.10(7) |